\documentstyle[epsfig,longtable]{aipproc}
\def\mic{$\mu$m\ }
\def\nwat{nW m$^{-2}$ sr$^{-1}$}
\begin{document}

\title{A Tentative Detection of the Cosmic Infrared Background at 3.5 \mic from {\it COBE}/DIRBE Observations}

\author{E. Dwek$^*$ and R. G. Arendt$^{\dagger}$}
\address{$^*$Laboratory for Astronomy and Solar Physics, Code 685,
NASA/Goddard
Space Flight Center, Greenbelt, MD 20771.\ \ e-mail address:\
eli.dwek@gsfc.nasa.gov\\
$^{\dagger}$Raytheon STX, Code 685, NASA/Goddard
Space Flight Center, Greenbelt, MD 20771}

\maketitle

\begin{abstract}
Foreground emission and scattered light from interplanetary dust (IPD)
particles and emission from
Galactic stellar sources are the
greatest obstacles for determining the cosmic infrared background (CIB)
from diffuse sky
measurements in the $\sim$ 1 to 5 \mic range. 
We describe a new method for deriving the CIB at near infrared
wavelengths which reduces the uncertainties associated with the
removal of the Galactic stellar emission component from the sky maps. 
The method produces positive residuals at 3.5 and 4.9 $\mu$m, of which
only the 3.5 \mic residual is nearly isotropic. We consider our result as a
tentative detection of the CIB at this wavelength.
\end{abstract}

\section*{Introduction}
Determination of the cosmic infrared background (CIB) from diffuse sky
measurements is
 greatly hampered by the presence of foreground emission and scattered
light from the
interplanetary dust (IPD) cloud, and emission 
from discrete and unresolved stellar components in our Galaxy, and from 
dust in the interstellar medium (ISM).
In a recent publication, Hauser et al. (1998; hereafter H98) presented
the results of the
search for the CIB in the 1.25 to 240 \mic wavelength region that was 
conducted with
the Diffuse Infrared Background Experiment (DIRBE)
 on the {\it Cosmic Background Explorer} ({\it COBE}) satellite. Careful
subtraction of foreground emission from the
IPD cloud
(Kelsall et al. 1998) and from stellar and interstellar Galactic emission
components (Arendt et al. 1998) revealed a residual
emission component in the DIRBE skymaps that, after detailed analysis of the
random and systematic uncertainties, was consistent with
 a positive signal at 100, 140, and 240~$\mu$m. Subsequent rigorous tests
showed that only the 140 and 240 \mic signals were
isotropic, a strict requirement for their extragalactic origin. Only upper
limits for the CIB intensity were given for
$\lambda$~=~1.25~$-$~60~$\mu$m, where the CIB
detection was hindered by residual emission from the IPD cloud. In the 1.25
to 4.9~\mic wavelength region,
 uncertainties in the subtraction of the Galactic stellar component
contributed to the uncertainties as well. The upper limits on the CIB
determined by H98 can be found in Table 2 of their paper.

Here we briefly summarize a new method for the subtraction of the
Galactic stellar emission component. Instead of using a statistical
 model to characterize the Galactic stellar emission (Arendt et
al. 1998), we create a spatial template of this emission from the
DIRBE data itself. The method is described in more detail by Dwek \& 
Arendt (1998; hereafter DA98). In this contribution we emphasize
isotropy tests conducted on the residual emission to examine its
possible extragalactic origin.

\section{Description of the Method}
\subsection{Subtraction of the Galactic Stellar Emission} 
We use the DIRBE 1.25, 2.2, 3.5, and 4.9 \mic all sky maps
from which the emission from interplanetary dust (IPD)
 has been subtracted (Kelsall et al. 1998) as the starting
point in the analysis. The
intensity of
 these maps should, in principle, contain only
the Galactic emission (starlight and ISM emission) and the CIB.

We then use the IPD$-$emission subtracted 2.2 \mic skymap to  create a
template map for the Galactic stellar emission component at this
wavelength.
The 2.2 \mic map is particularly suitable
for our analysis, since the contribution from
interstellar dust emission is negligible at this
wavelength. Furthermore, ground$-$based galaxy counts in the K-band
 (e.g. Gardner
1996) provide a strict
lower limit of 7.4 \nwat\ to the CIB intensity at this wavelength which
can be used as the nominal value for the CIB at 2.2
$\micron$. The  spatial template of the Galactic stellar emission
component is derived by subtracting this uniform intensity from the 2.2
$\micron$ zodi$-$subtracted skymap. The final results of our analysis
depend 
 only weakly on the adopted
2.2 $\micron$ CIB intensity, and we
explicitly state their dependence on this value.

The next step in the analysis consists of correlating the 2.2
\mic stellar
 emission template with
the zodi$-$subtracted skymaps at 1.25, 3.5, and 4.5 $\mu$m (see Figure 1
in DA98).
 
The slopes of the correlations represent the colors of stellar
emission, and are in very good agreement with the colors of M and K
giants shown in Figure 2 of Arendt et al. (1994).
The intercepts of the correlations represent the residual emission. At
3.5 and 4.9 \mic we had to subtract the small contribution of emission
from interstellar dust.

The final step in the analysis consisted of testing the residual
emission components for
positivity and isotropy, the two criteria required to ascertain their
extragalactic nature (H98). 
The analysis concentrated on high quality (HQ) regions of the sky, identified
as such for their location at high Galactic latitudes ($b$)
and ecliptic latitudes ($\beta$), in which the contributions from the 
Galactic and zodiacal
emission components are relatively small compared to
other regions of the sky. Figure 1 in H98 depicts the
location of these HQ regions.

The error analysis (see DA98 for further details) shows that our
method significantly reduces the errors associated with the
subtraction of Galactic starlight. However, only at 3.5 and 4.9 \mic
are the residuals positive, and as described in more detail below, 
only the 3.5 \mic residual is nearly isotropic. 
A panel of three DIRBE 3.5 \mic maps depicting the as observed
sky, the sky after the removal of the interplanetary dust emission,
and the final residual map after the subtraction of the Galactic
starlight and ISM emission was presented by Cowen (1998).

\subsection{Isotropy Tests}

Any extragalactic signal is expected to be isotropic, barring small
fluctuations expected from the discrete distribution of galaxies. We
therefore conducted several isotropy tests on the residual
intensities in the 3.5 and 4.9 \mic bands, which in increasing
order of complexity consisted of: (a) comparison of mean intensities in
the various HQ regions; (b) examination of large scale gradients; and
(c) analysis of the two point correlation function in the HQ regions.

Table 1 in DA98 shows that the residual 3.5~\mic emission is isotropic,
as evident from
the agreement between the mean intensities in the
various HQ regions. The 4.9 \mic residuals failed this simple isotropy
test.

Examination of large scale gradients in the residual 3.5 \mic map shows that
they are significantly reduced compared to those found in
the residual intensity map presented by H98. A
comparison between the cuts in the residual intensity maps derived by
the two methods is presented in Figure 1.

\begin{figure}[ht] 
\centerline{\epsfig{file=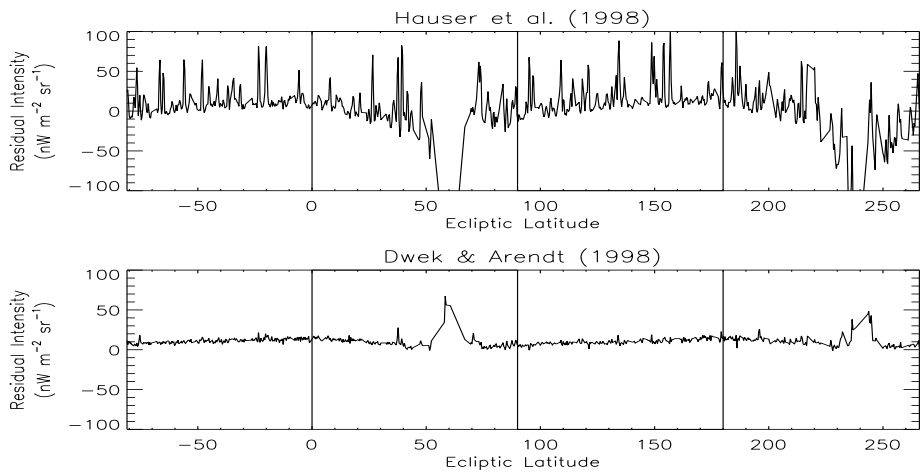,width=5.0in,height=2.5in}}
\vspace{6pt}
\caption{
(Top) Intensities in the Hauser et al. (1998) 3.5 $\mu$m residual map
along a great circle at ecliptic longitude 0\arcdeg and 180\arcdeg 
(the latter plotted as latitudes $>$ 90\arcdeg). All foreground models have 
been removed, and the locations of bright stars have been blanked.
(Bottom) The same intensity slice after the removal of the improved model
of the Galactic stellar emission (Dwek \& Arendt 1998).
}\label{theCutFigure}
\end{figure}

\begin{figure}[ht] 
\centerline{\epsfig{file=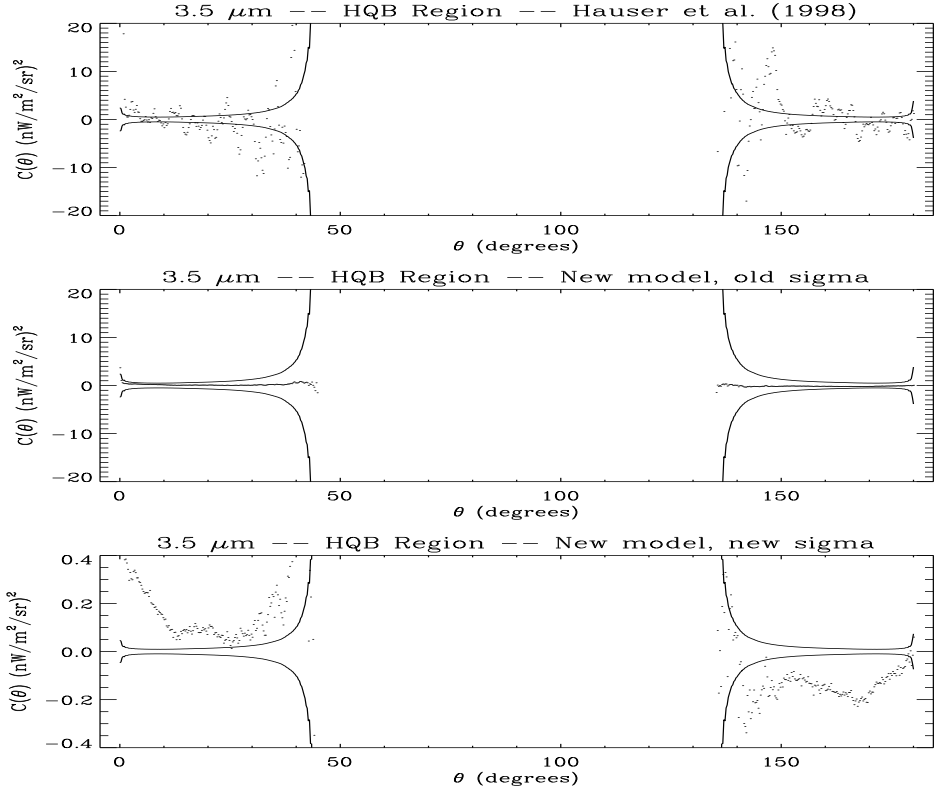,width=5.0in,height=4.20in}}
\vspace{8pt}
\caption{
Two-point correlation functions of the 3.5 $\mu$m residual emission in the
HQB region. (Top) The result after foreground subtractions as presented by
Hauser et al. (1998). (Middle) The result after foreground subtractions using
the improved model of the Galactic stellar emission (Dwek \& Arendt 1998)
shown with the same uncertainties as the Hauser et al. (1998) result. (Bottom)
The result after foreground subtractions from Dwek \& Arendt (1998), now
shown with uncertainties that do not include a component allowing for random
stellar variations. Note the large change in scale.
}\label{2ptCorrFigure}
\end{figure}

Although greatly reduced, the bottom panel of Fig. 1 still shows the
presence of large scale gradients in the residual map. At low Galactic
latitudes ($\beta \approx 60^{\circ},\ 240^{\circ}$ in Figure 1) 
the gradient is
largely due to residual Galactic stellar
emission which is affected by dust extinction. Dust extinction is
negligible in the HQ regions. Additional gradients are due to the
incomplete subtraction of the IPD emission.

The same gradient seen in the 3.5 \mic image is significantly 
stronger in the
4.9 \mic residual map, and is responsible for the larger
dispersion in the mean intensities in the various HQ regions. We
therefore consider the 4.9 \mic result as only an upper limit.

We also performed a two point correlation analysis of the residuals, a
strict isotropy test adopted by H98. In this test the
two point correlation function of the residual emission is compared to
that of a simulated flat background posessing random Gaussian
uncertainties estimated from the DIRBE data (H98). Figure 2 shows the 
results of our analysis.
 
The top panel shows the two point correlation function of the
residuals presented by H98 (Figure 3 in their paper). 
The middle panel shows the two point correlation function of the
residuals using our method for subtracting the Galactic stellar
emisison component, but with the same uncertainties in the residual
maps as those adopted by H98. These uncertainties are dominated by
unremoved stellar emission, visible as small$-$scale structure in the
residuals shown in Figure 1. With
these uncertainties, the residual emission is isotropic. The bottom
panel 
shows the two point correlation function of the residuals obtained by
using the new method for foreground subtraction, and their reduced
uncertainties. Note the large change in scale compared to the previous
panels. Because of the
smaller uncertainties, the residual emission does not pass the strict
two point correlation isotropy test.

If we used only the the mean intensities in the various HQ patches as
the criterion for isotropy, the 3.5 \mic residual would classify as a
definite detection of the CIB. However, because of the persistence of
large scale gradients (albeit largely reduced compared to the previous
analysis), and the existence of correlations between the pixel
intensities beyond what one would expect from a flat background, we
only regard the derived residual intensity as
a tentative detection of the CIB at 3.5 \mic with an intensity given by:

\begin{equation}
\nu I_{res} ({\rm nW\ m}^{-2} {\rm\ sr}^{-1})  =    
                   9.9 + 0.312\ \Delta_{CIB}(2.2~\mu \rm m) \pm 2.9 
\end{equation}
\noindent
where the quoted 
errors represent 1 $\sigma$ uncertainties, and 
$\Delta_{CIB}(2.2~\mu \rm m) \equiv [I_{CIB}(2.2\ \mu{\rm m}) - 7.4]$
represents the difference in the actual value of the CIB at
2.2~$\micron$ 
(in \nwat) and the nominal value adopted in our model. Our analysis
also yields new upper limits (95\% confidence limit) on the CIB at
1.25 and 4.9 \mic of 68 and 36 \nwat, respectively.

\end{document}